\pdfoutput=1
\documentclass[lettersize,journal]{IEEEtran}
\usepackage{amsmath,amsfonts}
\usepackage{algorithmic}
\usepackage{algorithm}
\usepackage{array}
\usepackage[caption=false,font=normalsize,labelfont=sf,textfont=sf]{subfig}
\usepackage{textcomp}
\usepackage{stfloats}
\usepackage{url}
\usepackage{verbatim}
\usepackage{graphicx}
\usepackage{cite}
\usepackage{comment}
\usepackage{amssymb}
\usepackage[utf8]{inputenc}
\usepackage{lipsum}
\usepackage{kantlipsum}
\usepackage[caption=false,font=normalsize,labelfont=sf,textfont=sf]{subfig}
\usepackage{graphicx}
\usepackage{float}
\usepackage{multirow}
\usepackage{booktabs}
\usepackage{changepage}

\usepackage{tikz}
\usetikzlibrary{quantikz2}

\begin{document}

\title{Dynamic Runtime Assertions in Quantum Ternary Systems}

\author{Ehsan Faghih, \and Huiyang Zhou
\thanks{E. Faghih is with the Department of Electrical and Computer Engineering, NC State University, Raleigh, NC, US. sfaghih@ncsu.edu}
\thanks{H. Zhou is with the Faculty of Department of Electrical and Computer Engineering, NC State University, Raleigh, NC, US. hzhou@ncsu.edu}
}

\maketitle

\begin{abstract}
With the rapid advancement of quantum computing technology, there is a growing need for new debugging tools for quantum programs. Recent research has highlighted the potential of assertions for debugging quantum programs. In this paper, we investigate assertions in quantum ternary systems, which are more challenging than those in quantum binary systems due to the complexity of ternary logic.  
We propose quantum ternary circuit designs to assert classical, entanglement, and superposition states, specifically geared toward debugging quantum ternary programs.
\end{abstract}


\section{Introduction}
Quantum computing has distinctive advantages compared to classical computing, and the latest breakthroughs in quantum computer hardware have ignited optimistic prospects for unlocking the extraordinary potential of this field. Furthermore, it is shown that realizing quantum computing structures using multi-valued logic can bring many advantages over its binary counterpart \cite{de2009multiple},\cite{muthukrishnan2000multivalued}. A careful examination of the product of the number's width (amount of digits) and the depth of digit (maximum number of symbols in each digit) as an influential factor of hardware cost in digital systems showed that the most economical radix is three.  Quantum systems are no exception \cite{khan2007quantum},\cite{jc2003qutrit}. Recent studies have shown that the design of quantum computers based on qutrit, the unit of quantum information in ternary representation, brings 37\% more compactness than the quantum computer based on qubit, the unit of quantum information in binary representation \cite{FAGHIH2023100908}, \cite{haghparast2017towards}. Quantum Ternary logic finds practical applications in the development of ternary computers such as quantum multiple-valued decision diagrams (QMDD) \cite{miller2022multiple}. Ternary logic outperforms binary logic in several ways, one of which is its capacity to convey more information using fewer digits. This advantage enhances the flexibility for encoding and processing data. Additionally, it simplifies circuitry by reducing the necessity for numerous gates and connections, ultimately leading to reduced energy consumption \cite{miller2022multiple}. It was also shown recently that quantum circuits via qutrits can introduce asymptotic improvements~\cite{Gokhale_2019}.
In quantum mechanics, ternary circuits can be physically realized with various technologies such as ion-trap \cite{de2009multiple}. 

The development of quantum computing systems at a large scale, enabling the execution of algorithms on extensive datasets, calls for better tools. Assertions are a primitive that can be used for both program debugging \cite{huang2018qdb}, \cite{liu2020quantum} and error mitigation \cite{Li_2022}. 
However, these prior works on quantum assertion were developed for quantum circuits with qubits. 
To further enhance the concept and leverage the benefits of quantum ternary logic, this paper focuses on supporting assertions in quantum circuits with ternary bits or qutrits.

The remainder of the paper is organized as follows. Section \ref{sec:Background} provides background on quantum ternary gates and their operations and summarizes the prior works on quantum assertion. In Section \ref{sec:quan-circuits}, we delve into our proposed designs for dynamic assertions in quantum ternary circuits, elaborating on their functioning. Section \ref{sec:Evaluation} comprises an evaluation utilizing predefined test cases. Finally, Section \ref{sec:Conclusion} concludes.

\section{Background}
\label{sec:Background}
\subsection{Quantum Ternary Logic}
Quantum ternary-valued logic processors are a class of quantum systems in which each information unit, referred to as a qutrit, can be represented using three distinct 3×1 matrices. The states \(|0\rangle\), \(|1\rangle\), and \(|2\rangle\) are considered the fundamental states, aka the computational basis states, of a qutrit, each possessing a distinctive representation as follows:
\begin{equation}\label{eq1}
|0\rangle = \begin{bmatrix} 1 \\ 0 \\ 0 \end{bmatrix} \quad |1\rangle = \begin{bmatrix} 0 \\ 1 \\ 0 \end{bmatrix} \quad |2\rangle = \begin{bmatrix} 0 \\ 0 \\ 1 \end{bmatrix}
\end{equation} \newline
When considering complex numbers \(\alpha \), \(\beta \), and \(\gamma \), a qutrit can exist in a superposition state, simultaneously occupying a linear combination of the computational basis states \(|0\rangle\), \(|1\rangle\), and \(|2\rangle\). This superposition state is denoted by \(|\psi \rangle\) = \(\alpha \)\(|0\rangle\) + \(\beta \)\(|1\rangle\) + \(\gamma \)\(|2\rangle\), where \(\lvert\alpha\rvert^2 \) + \(\lvert\beta\rvert^2 \) + \(\lvert\gamma\rvert^2 \) = 1. For an n-qutrit system, there are \(3^n\) different computational basis states. For instance, in a two-qutrit system, the state can be represented as: \(|\psi \rangle\) = \(\alpha_{00}\)\(|00\rangle\) + \(\alpha_{01}\)\(|01\rangle\) + \(\alpha_{02}\)\(|02\rangle\) + \(\alpha_{10}\)\(|10\rangle\) + \(\alpha_{11}\)\(|11\rangle\) + \(\alpha_{12}\)\(|12\rangle\) + \(\alpha_{20}\)\(|20\rangle\) + \(\alpha_{21}\)\(|21\rangle\) + \(\alpha_{22}\)\(|22\rangle\), where \(\alpha\) is a complex coefficient, and \(\sum_{\delta\in\{0,1,2\}^2}\lvert\alpha\rvert^2 = 1\).

There are six single-qutrit [Z] gates, and each of them is associated with a corresponding unitary 3×3 matrix, as illustrated below. Their functionalities are shown in Table 1.
\[Z(0) = \begin{bmatrix} 1&0&0\\ 0&1&0 \\ 0&0&1 \end{bmatrix} \hspace{0.6em} Z(+1) = \begin{bmatrix} 0&0&1\\ 1&0&0 \\ 0&1&0 \end{bmatrix} \hspace{0.6em} Z(+2) = \begin{bmatrix} 0&1&0\\ 0&0&1 \\ 1&0&0 \end{bmatrix}\] 
\[Z(01) = \begin{bmatrix} 0&1&0\\ 1&0&0 \\ 0&0&1 \end{bmatrix} \hspace{0.6em} Z(02) = \begin{bmatrix} 0&0&1\\ 0&1&0 \\ 1&0&0 \end{bmatrix} \hspace{0.6em} Z(12) = \begin{bmatrix} 1&0&0\\ 0&0&1 \\ 0&1&0 \end{bmatrix}\]

In the context of qutrit systems, it is widely accepted that the quantum cost associated with single qutrit gates (like [Z] gates) and a controlled-[Z] gates is considered to be equal to unity \cite{mohammadi2009figures} It is because, as far as our knowledge extends, there is currently no established benchmark for ternary quantum systems. In this study, we adopt the convention of assigning a quantum cost of unity to each M-S gate and Chrestenson gate. 

\vspace{0.4cm}
\textbf{Ternary Muthukrishnan-Stroud (M-S) Gates:} A M-S gate consists of two types of primitive ternary quantum gates, namely 1-qutrit and 2-qutrit gates. 
Single-qutrit gates operate based on the Z transforms mentioned before, whereas two-qutrit gates include a control input for performing the Z transform, i.e., Controlled-[Z] gates. It means that only when the control qutrit is \(|2\rangle\), the gate triggers the Z transformation on the target qutrit. 

\begin{table}
\begin{center}
\caption{Permutations of 1-qutrit M-S gates}
  \label{tab:tbl1}
  
  \centering
  \small 
  \noindent\begin{tabular}{| c | c | c | c | c | c | c |}
    \hline
    \multicolumn{7}{|c|}{\textbf{PERMUTATIONS}} \\
    \hline
    \textbf{Input} & Z(0) & Z(+1) & Z(+2) & Z(12) & Z(01) & Z(02) \\
    \hline
    \textbf{0} &0&1&2&0&1&2 \\
    \textbf{1} &1&2&0&2&0&1 \\
    \textbf{2} &2&0&1&1&2&0 \\
    \hline
  \end{tabular}
  \end{center}
\end{table}
\vspace{0.4cm}
\textbf{Chrestenson basis:} Analogous to the Hadamard basis, i.e., \(|+\rangle\) and \(|-\rangle\) states, in qubit systems, 
 
the Chrestenson basis \cite{Hurst1985-HURSTI-2} serves as a natural extension to the Hadamard basis in qutrit systems. 
There are two Chrestenson gates known as Ch1 and Ch2 which correspond to Hadamard gates in qubit systems. It is important to note that \(Ch1Ch2 = I\) (identity matrix). Here, $\omega$ is the cube root of unity, i.e., $\omega^3 = 1$, and $1 + \omega + \omega^2 = 0$ \cite{gottesman1998fault}.
\begin{align*}
    |+\rangle &= \quad\frac{1}{\sqrt{3}} (|0\rangle + |1\rangle + |2\rangle) \\
    |-_{i}\rangle &= \quad\frac{1}{\sqrt{3}} (|0\rangle + \omega^i|1\rangle + \omega^{2i}|2\rangle) \quad i\in\{1,2\} \\
    ||_{i}\rangle &= \quad\frac{1}{\sqrt{3}} (|0\rangle + \omega^{2i}|1\rangle + \omega^i|2\rangle) \quad i\in\{1,2\}
\end{align*}
\[Ch1 = \begin{bmatrix} 1&1&1\\ 1&\omega&\omega^2 \\ 1&\omega^2&\omega \end{bmatrix} \hspace{0.6em} Ch2 = \begin{bmatrix} 1&1&1\\ 1&\omega^2&\omega \\ 1&\omega&\omega^2 \end{bmatrix}\] 

\subsection{Quantum Assertions}

Huang et al. \cite{huang2019statistical} proposed a statistical approach for quantum assertion. 
They identified three essential types of assertions for debugging quantum programs: classical assertions, superposition assertions, and entanglement assertions. 
Classical assertions involve checking quantum variables against classical values to determine if they match the desired values.
Superposition assertions are used to verify whether a quantum variable is in a desired superposition state.
Entanglement assertions focus on checking whether the entangled quantum variables exhibit the desired correlation. Statistical assertions require measurements of the qubits of interest, thereby being disruptive to program execution.

Liu et al. introduced the concept of dynamic quantum assertions \cite{liu2020quantum}, which means that the assertion check is performed during program execution and the program continues execution if there is no assertion error. In this paper, we introduce dynamic assertion circuits for classical, superposition, and entangled states in the context of ternary logic.

Liu et al. further proposed two systematic approaches for dynamic quantum state assertion, capable of asserting a broader range of quantum states, including pure and mixed states \cite{liu2021systematic}.

Li et al. introduced Proq, a runtime assertion scheme for quantum program testing and debugging \cite{li2020projection}. Proq utilizes projections based on Birkhoff-von Neumann quantum logic, enabling efficient assertion verification through minimal measurements. Their work demonstrates the efficacy of projection-based assertions for bug detection and ensuring program semantics in both exact and approximate quantum programs.

Enabling runtime assertion in ternary quantum circuits shares similar challenges to those found in binary quantum computing and is demanding for two primary reasons. Firstly, the non-cloning theorem poses a fundamental limitation by prohibiting the exact replication of qubits, making conventional debugging and assertion checks challenging. Secondly, measuring a qubit results in the collapse of its superposition state into a classical state, leading to the loss of inherent parallel information. This unique challenge persists in the verification of assertions within quantum states. 
Additionally, in the context of ternary quantum circuits, there is a notable deficiency of strategies for error correction or bug detection. 
One potential reason is that the additional quantum state introduces more opportunities for errors. Designing robust error correction or detection systems that work effectively with ternary logic is an area of active research. Furthermore, designing and optimizing quantum gates for ternary logic is more complex compared to binary gates. Ternary gates, such as Toffoli gates with ternary inputs, have to account for three quantum states, making gate design and optimization more intricate \cite{khan2007quantum}.

\section{Quantum Ternary Circuits For Dynamic Assertions}
\label{sec:quan-circuits}
Our approach to enabling dynamic assertions revolves around the introduction of additional quantum bits, referred to as ancilla qutrits. These ancilla qutrits provide information about the qutrits under test. Instead of directly measuring the qutrits under test, we measure the ancilla qutrits. This allows us to verify assertions without disrupting the program execution. However, it is crucial to ensure that measuring the ancilla qutrits does not impact the original quantum circuit. In the following sections, we outline our proposed circuits for each type of assertion. 

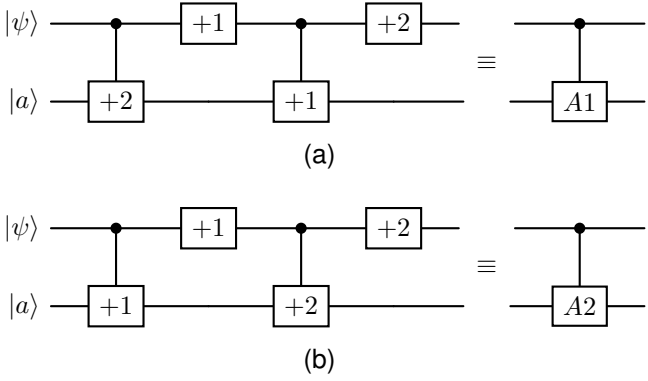
\begin{figure}[htbp]
  \centering
  \subfloat[]{
\label{fig:a1}%
    \begin{quantikz}
      \lstick{$\ket{\psi}$} & \ctrl{1} & \gate{+1} & \ctrl{1} & \gate{+2} & \midstick[2,brackets=none]{\(\equiv\)} & \ctrl{1}& \\
      \lstick{$\ket{a}$}    & \gate{+2} &           & \gate{+1} &            &                                & \gate{A1} &
    \end{quantikz}}%

  \subfloat[]{
\label{fig:b1}%
    \begin{quantikz}
      \lstick{$\ket{\psi}$} & \ctrl{1} & \gate{+1} & \ctrl{1} & \gate{+2} & \midstick[2,brackets=none]{\(\equiv\)} & \ctrl{1}& \\
      \lstick{$\ket{a}$}    & \gate{+1} &           & \gate{+2} &            &                                & \gate{A2} &
    \end{quantikz}}%
  \caption{The A1 and A2 circuits and their gate scheme symbol. \(\ket{a}\) is an ancilla input.}
  \label{fig:A1A2}
\end{figure}

To develop these assertion circuits, we first design two simple ternary circuits called A1 and A2 gates, depicted in Fig. \ref{fig:A1A2}.

An A1 gate adds its control qutrit value to its target qutrit. 
In the figure, the target is $\ket{a}$, and the control qutrit is $\ket{\psi}$. For example, if the control qutrit is set to 1 and the target is 0, the target would become 1 after A1, and the control qutrit remains the same. Similarly, A2 employs a similar principle to add (2 *  controller value) to the target. Using the same example, if the controller is also set to 1, it will add 2 to its target. The gates A1 and A2 are formally defined as follows. Note that these gates become identity gates when the control bit is 0, and \(i\in\{1,2\}\).
\begin{equation*}\label{eq2:A1A2}
\begin{aligned}
\begin{cases}
    A1 : \text{if }(control=i) \text{ then target} = |\hspace{0.05cm}target+i\rangle \hspace{0.1cm}mod \hspace{0.1cm}3.\\ A2 : \text{if }(control=i) \text{ then target} = |\hspace{0.05cm}target+2i\rangle\hspace{0.1cm}mod \hspace{0.1cm}3. 
\end{cases}
\end{aligned}
\end{equation*}
As shown in Fig. \ref{fig:A1A2}, the implementation of the gates involves the use of two 1-qutrit and two 2-qutrit gates. Consequently, the quantum cost of these operations is equal to 4, and their depth is also 4.

\subsection{Dynamic Assertion for Ternary Classical Values}
\label{subsec:Classical-Asr}
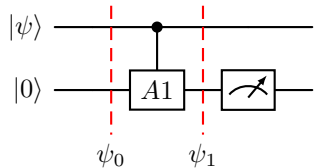
\begin{figure}[htbp]
  \centering
      \begin{quantikz}[
slice all, remove end slices=1,
slice style = {shorten <=-0.1cm,
               shorten >=-0.1cm},
slice label style = {yshift=-22mm}, 
slice titles = $\psi_{\fpeval{\col-1}}$
                    ]
        \lstick{$\ket{\psi}$}& &  \ctrl{1} && \\
        \lstick{$\ket{0}$}    &  &\gate{A1}& \meter{} & 
      \end{quantikz}
  \caption{Circuit for classical-value assertion, assert(\(\ket{\psi}\) == \(\ket{0}\)).}
  \label{fig:classicAsr}
\end{figure}

By asserting for ternary classical values, we aim to assert (\(\ket{\psi}\) == \(\ket{i}\)), where \(\ket{\psi}\) is the qutrit of interest and \(i\in\{0,1,2\}\). To achieve this, we propose the circuit depicted in Fig. \ref{fig:classicAsr}. 
The circuit shown in Fig. \ref{fig:classicAsr} checks whether the state of \(\ket{\psi}\) is equal to \(\ket{0}\). The ancilla qutrit is initially set to \(\ket{0}\) and is measured after the A1 gate operation. By initializing the ancilla qutrits to \(\ket{i}\), the same circuit can be used to assert (\(\ket{\psi}\) ==\(\ket{2i\hspace{0.1cm} mod\hspace{0.1cm} 3}\)).

\textit{\textbf{Proof.}} Let us consider the case where the ancilla input is set to \(\ket{0}\), and we are asserting whether \(\ket{\psi}\) is equal to \(\ket{0}\). If \(\ket{\psi}\) is in a classical state, taking values of either \(\ket{0}\), \(\ket{1}\), or \(\ket{2}\), the resulting states of the \(\ket{\psi_0}\) can be represented as \(\ket{00}\), \(\ket{10}\), or \(\ket{20}\). Therefore,
the resulting states, denoted as \(\ket{\psi_1}\) , would be \(\ket{00}\), \(\ket{11}\), or \(\ket{22}\), respectively. Consequently, when the ancilla qutrit is measured and yields the state \(\ket{0}\), it signifies that \(\ket{\psi}\) must be \(\ket{0}\), indicating no assertion error. If the measurement outcome is \(\ket{1}\), it implies that \(\ket{\psi}\) must be \(\ket{1}\), and if it is \(\ket{2}\), \(\ket{\psi}\) must be \(\ket{2}\), indicating an assertion error. 

If the \(\ket{\psi}\) is in a superposition state, represented as \(\ket{\psi}\) = a\(\ket{0}\) + b\(\ket{1}\) + c\(\ket{2}\) due to a bug or runtime error, the resulting \(\ket{\psi_0}\) becomes a\(\ket{00}\) + b\(\ket{10}\) + c\(\ket{20}\) and \(\ket{\psi_1}\) becomes a\(\ket{00}\) + b\(\ket{11}\) + c\(\ket{22}\), indicating an entangled state. This entanglement leads to a unique behavior during the measurement of the ancilla qutrit. If the measurement outcome is \(\ket{0}\) (no assertion error), the qutrit under test is projected into the classical state \(\ket{0}\), denoted as \(\ket{\psi'}\) = \(\ket{0}\). Conversely, if the measurement outcome is \(\ket{1}\) (an assertion error), it is projected into the classical state \(\ket{1}\). In the context of an assertion check (\(\ket{\psi}\) == \(\ket{0}\)), the proposed circuit has the potential to automatically correct the qutrit if it is in a superposition state, resulting in no assertion error. However, if the qutrit cannot be corrected into the expected classical state, an assertion error occurs. The probability of obtaining a measurement result of \(\ket{0}\), \(\ket{1}\) or \(\ket{2}\) is determined by the squared magnitudes of coefficients \(|a|^2\), \(|b|^2\) and \(|c|^2\), respectively. The probability distribution of
assertion errors over multiple runs can be used to estimate
a, b, and c, as needed. 

The cases when the ancilla qutrit is set to \(\ket{1}\) and \(\ket{2}\) can be derived similarly to assert for \(\ket{2}\) and \(\ket{1}\), respectively.

\subsection{Dynamic Assertion for Entangled States}
\label{subsec:entangled-Asr}
We propose dedicated circuits to check whether two qutrits under test are in certain entangled states. Fig \ref{fig:entangled_groups} shows the two groups of commonly used entangled states, and we propose two circuits as shown in Fig \ref{fig:entangled_circuits} to assert them, respectively. 

\begin{figure}[htbp]
  \centering
\begin{tikzpicture}
  \node [draw, rectangle, align=center] (box) at (0, 0) {%
  a\(\ket{00}\) + b\(\ket{12}\) + c\(\ket{21}\) \\ a\(\ket{01}\) + b\(\ket{10}\) + c\(\ket{22}\)\\
    a\(\ket{02}\) + b\(\ket{11}\) + c\(\ket{20}\)
  };
  \node [below] at (box.south) {(a)};
      
\end{tikzpicture}
\begin{tikzpicture}
  \hspace{0.5cm}\node [draw, rectangle, align=center] (box) at (0, 0) {%
  a\(\ket{00}\) + b\(\ket{11}\) + c\(\ket{22}\) \\ a\(\ket{02}\) + b\(\ket{10}\) + c\(\ket{21}\)\\
    a\(\ket{01}\) + b\(\ket{12}\) + c\(\ket{20}\)
  };
  \node [below] at (box.south) {(b)};
\end{tikzpicture}
\caption{Two entangled groups of ternary states}
  \label{fig:entangled_groups}
\end{figure}
In Fig. \ref{fig:entangled_circuits}, circuit (a) is designed for asserting entangled states in group (a) of Fig. \ref{fig:entangled_groups}. By setting the ancilla qutrits to different initial states, this circuit verifies whether the qutrit of interest \(\ket{\psi_{0}}\) is in one of the entangled states listed in group (a). Next, we explain the assertion process for asserting the ternary state a\(\ket{00}\) + b\(\ket{12}\) + c\(\ket{21}\), when the ancilla bit is set to \(\ket{0}\)).  

\textit{\textbf{Proof.}} 
\(\ket{\psi_{0}}\) = a\(\ket{000}\) + b\(\ket{120}\) + c\(\ket{210}\) \\
\(\ket{\psi_{1}}\) = a\(\ket{000}\) + b\(\ket{121}\) + c\(\ket{212}\) \\
\(\ket{\psi_{2}}\) = a\(\ket{000}\) + b\(\ket{120}\) + c\(\ket{210}\) = \(\ket{\psi}\otimes\) \(\ket{0}\)\\

\(\ket{\psi}\otimes\) \(\ket{0}\) is the ternary entangled state that we intended to assert, along with an un-entangled ancilla qutrit. By measuring the ancilla bit, we can determine if we have successfully achieved the desired state without collapsing the entangled state. If the measurement yields a zero, it indicates that the system is in the correct state and the entanglement is preserved.

\begin{figure}[H]
  \centering
  \subfloat[]{
  \label{fig:3a}%
  
    \begin{quantikz}[
        slice all, remove end slices=1,
        slice style = {shorten <=-0.1cm,
                       shorten >=-0.1cm},
        slice label style = {yshift=-28mm}, 
        slice titles = $\psi_{\fpeval{\col-1}}$
        ]
      \lstick[2]{$\ket{\psi}$} && \ctrl{2} & & &  \\ 
      && & \ctrl{1} &   &         \\
      \lstick{$\ket{a}$}& & \gate{A1} & \gate{A1} &\meter{}& 
    \end{quantikz}}
    
  \subfloat[]{
   \label{fig:3b}%
        \begin{quantikz}
        [
        slice all, remove end slices=1,
        slice style = {shorten <=-0.1cm,
                       shorten >=-0.1cm},
        slice label style = {yshift=-28mm}, 
        slice titles = $\psi_{\fpeval{\col-1}}$
        ]
      \lstick[2]{$\ket{\psi}$} && \ctrl{2} & & &  \\ 
      && & \ctrl{1} &   &         \\
      \lstick{$\ket{a}$}& & \gate{A1} & \gate{A2} &\meter{}& 
    \end{quantikz}}
  \caption{Two proposed circuits to assert ternary entanglement. (a) Proposed circuit to assert the entangled state in group \textit{a} of Fig \ref{fig:entangled_groups}. (b) Proposed circuit to assert the entangled state in group \textit{b} of Fig \ref{fig:entangled_groups}.}
  \label{fig:entangled_circuits}
\end{figure}
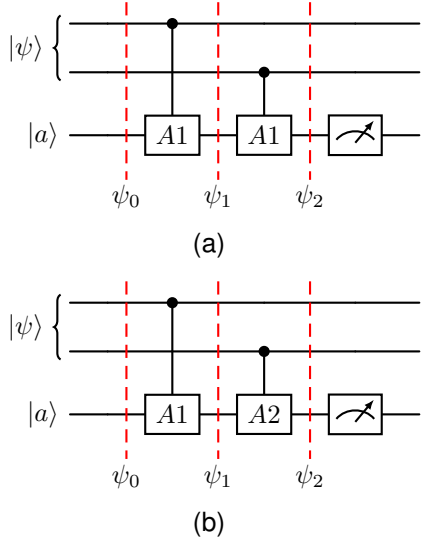
If the input qutrits are not entangled in the expected state, it can be expressed as \(\ket{\psi}\) = a\(\ket{00}\)+ d\(\ket{01}\)+ g\(\ket{02}\)+ e\(\ket{10}\)+ h\(\ket{11}\)+ b\(\ket{12}\)+ i\(\ket{20}\)+ c\(\ket{21}\)+ f\(\ket{22}\). Then, the circuit produces the following states:
\begin{small}
\begin{align*}
\ket{\psi_{0}} &= [a\ket{000} + d\ket{010} + g\ket{020} 
+ e\ket{100} + h\ket{110} + b\ket{120}
+ \\ &+ i\ket{200} + c\ket{210} + f\ket{220}].\\
\ket{\psi_{1}} &= [a\ket{000} + d\ket{010} + g\ket{020} 
+ e\ket{101} + h\ket{111} + b\ket{121}
+ \\ &+ i\ket{202} + c\ket{212} + f\ket{222}].\\
\ket{\psi_{2}} &= [a\ket{000} + d\ket{011} + g\ket{022} 
+ e\ket{101} + h\ket{112} + b\ket{120}
+ \\ &+ i\ket{202} + c\ket{210} + f\ket{221}]. 
\end{align*}
\end{small}

\begin{table}[htbp]
  \begin{center}
  \caption{Ternary circuit outputs based on corresponding ancilla and $\ket{\psi}$ states, referring to Fig. \ref{fig:entangled_circuits}. and Fig. \ref{fig:entangled_groups}.}
  \label{tab:tblx}

  \centering
  \small 
  \noindent\begin{tabular}{| c | c | c | c |}
    \hline
    Circuit &$\ket{\psi}$ & Ancilla value & Output  \\ 
    \hline
    \multirow{3}{*}{\rotatebox{90}{\begin{small} Fig.\ref{fig:3b}
    \end{small}}}&$a\ket{00}+b\ket{11}+c\ket{22}$ & $\ket{0}$   & $\ket{\psi}\otimes\ket{0}$  \\
    &$a\ket{02}+b\ket{10}+c\ket{21}$ & $\ket{2}$   & $\ket{\psi}\otimes\ket{0}$  \\
    &$a\ket{01}+b\ket{12}+c\ket{20}$ & $\ket{1}$   & $\ket{\psi}\otimes\ket{0}$  \\
    \hline
    \multirow{3}{*}{\rotatebox{90}{\begin{small} Fig.\ref{fig:3a}
    \end{small}}}&$a\ket{00}+b\ket{12}+c\ket{21}$ & $\ket{0}$   & $\ket{\psi}\otimes\ket{0}$  \\
    &$a\ket{01}+b\ket{10}+c\ket{22}$ & $\ket{2}$   & $\ket{\psi}\otimes\ket{0}$  \\
    &$a\ket{02}+b\ket{11}+c\ket{20}$ & $\ket{1}$   & $\ket{\psi}\otimes\ket{0}$  \\
    
    \hline
  \end{tabular}
  
  \end{center}
\end{table}
When measuring the ancilla qutrit, the result can be either \(\ket{0}\), \(\ket{1}\) or \(\ket{2}\). If the result is \(\ket{0}\), the state \(\ket{\psi_{2}}\) is projected to a\(\ket{000}\) + b\(\ket{120}\) + c\(\ket{210}\) = (a\(\ket{00}\) + b\(\ket{12}\) + c\(\ket{21})\otimes\ket{0}\), forcing the input qutrits into an entangled state. Likewise, if the result is \(\ket{1}\) or \(\ket{2}\), the state \(\ket{\psi_{2}}\) is projected to the $(d\ket{011} + e\ket{101} + f\ket{221})$ or $(g\ket{022} + h\ket{112} + i\ket{202})$ terms, respectively representing different entangled states. In these cases, an assertion error would be reported as the measurement result is not $0$. The probability of measuring \(\ket{0}\), \(\ket{1}\) or \(\ket{2}\) can be used to compute the coefficients \textit{a} to \textit{i}, if needed.

Following similar steps, we can see that by initializing the ancilla qutrits to \(\ket{1}\) or \(\ket{2}\), the same circuit can be employed to assert whether the state  \(\ket{\psi}\) is equal to a\(\ket{02}\) + b\(\ket{11}\) + c\(\ket{20}\) or a\(\ket{01}\) + b\(\ket{10}\) + c\(\ket{22}\) respectively. This allows for the reuse of the circuit with different initializations of the ancilla qutrits to verify different desired states. 

The same proof can also be applied to assert the entangled states within the group $b$. Table \ref{tab:tblx} lists the states to be asserted along with the proper ancilla qutrit settings and the circuit to be used.

\subsection{Dynamic Assertion for Superposition}
\label{subsec:Superposition-Asr}

\begin{figure}[htbp]
  \centering
      \begin{quantikz}[
slice all, remove end slices=1,
slice style = {shorten <=-0.1cm,
               shorten >=-0.1cm},
slice label style = {yshift=-26mm}, 
slice titles = $\psi_{\fpeval{\col-1}}$
        ]
        \lstick{$\ket{\psi}$}& &  & \gate{A1}& && \\
        \lstick{$\ket{0}$}    &  &\gate{Ch1}& \ctrl{-1} & \gate{Ch2} &\meter{} & 
      \end{quantikz}
  \caption{Circuit for asserting equal superposition.}
  \label{fig:superpositionAsr}
\end{figure}
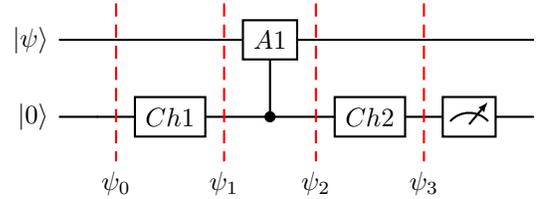

In binary quantum computing, a common pattern is to use Hadamard gates to put input qubits into an equal/uniform superposition state, denoted as \(\ket{+}\) = \(\tfrac{1}{\sqrt{2}} (\ket{0} + \ket{1}) \).  
As mentioned earlier, in ternary quantum computing, Ch1 and Ch2 gates, which operate based on the Chrestenson basis, serve a similar purpose. To verify such uniform superposition states, including \(\ket{+}\), \(\ket{-_{1}}\) or \(\ket{-_{2}}\), we propose a circuit as shown in Fig. \ref{fig:superpositionAsr}. 

\textit{\textbf{Proof.}} As shown in Fig \ref{fig:superpositionAsr}, if \(\ket{\psi}\) is equal to \(\ket{+}\) and the ancilla is initialized to \(\ket{0}\), the circuit produces the following states:

\begin{small}
\begin{align*}
\ket{\psi_{0}} &= a\ket{00} + b\ket{10} + c\ket{20} , \hspace{0.7mm}(a=b=c= \tfrac{1}{\sqrt{3}})\\
\ket{\psi_{1}} &= \tfrac{1}{3}[ \ket{0}\otimes(\ket{0}+\ket{1}+\ket{2}) +\\
&+\ket{1}\otimes(\ket{0}+\ket{1}+\ket{2}) + \\
&+\ket{2}\otimes(\ket{0}+\ket{1}+\ket{2})]\\
&= \tfrac{1}{3}[(\ket{00}+\ket{01}+\ket{02}) +\\
&+(\ket{10}+\ket{11}+\ket{12}) +\\
&+(\ket{20}+\ket{21}+\ket{22}) ]\\
\ket{\psi_{2}} &= \tfrac{1}{3} [\ket{00} + \ket{11} + \ket{22} + \ket{10} + \ket{21} + \ket{02} + \ket{20} + \ket{01} + \ket{12}] \\
\ket{\psi_{3}} &= \tfrac{1}{3\sqrt{3}}[\ket{00} + \ket{01} + \ket{02} + \ket{10} + \omega^2\ket{11} + \omega\ket{12} + \ket{20} \\ &+ \omega\ket{21} + \omega^2\ket{22} + \\
&+ \ket{10} + \ket{11} + \ket{12} + \ket{20} + \omega^2\ket{21} + \omega\ket{22} + \ket{00} +\\ &+ \omega\ket{01} + \omega^2\ket{02} + \\
&+ \ket{20} + \ket{21} + \ket{22} + \ket{00} + \omega^2\ket{01} + \omega\ket{02} + \ket{10} +\\ &+ \omega\ket{11} + \omega^2\ket{12} ]\\
&=> \\
\ket{\psi_{3}} &= \tfrac{3}{3\sqrt{3}} [\ket{00} + \ket{10} + \ket{20}] = \ket{+}\otimes\ket{0}
\end{align*}
\end{small}

Given that $(\omega^3 = 1)$ and $(\omega^2 + \omega + 1 = 0)$, and if the qutrit is in the uniform superposition state, denoted as $|\psi\rangle = |+\rangle$, then the coefficients $a$, $b$ and $c$ are equal to $1/\sqrt{3}$. In this case, $\ket{\psi_{2}}$ would be as described in the above-mentioned Proof. The expression of $\ket{\psi_{2}}$ clearly indicates that the two qutrits in the circuit are entangled. To resolve this entanglement, an additional $Ch2$ gate is included at the end of the circuit. Consequently, the state $\ket{\psi_{3}}$ will be equal to $\tfrac{1}{\sqrt{3}} \hspace{0.2mm}[\ket{00} + \ket{10} + \ket{20}] = \ket{+} \otimes \ket{0}$. We can follow the same process to assert for other uniform superposition states as shown in Table \ref{tab:tbl2}.

\begin{table}[htbp]
\begin{center}
  \caption{Output Analysis of the Circuit in Fig. \ref{fig:superpositionAsr} for Various Ancilla and $\ket{\psi}$ States.}
  \label{tab:tbl2}

  \centering
  \small 
  \noindent\begin{tabular}{| c | c | c |}
    \hline
    $\ket{\psi}$ & Ancilla value & Output  \\ 
    \hline
    $\ket{+}$ & $\ket{0}$   & $\ket{+}\otimes\ket{0}$  \\
    $\ket{+}$ & $\ket{1}$   & $\ket{+}\otimes\ket{1}$  \\
    $\ket{+}$ & $\ket{2}$   & $\ket{+}\otimes\ket{2}$  \\
    \hline
    $\ket{-_{1}}$ & $\ket{0}$   & $\ket{-_{1}}\otimes\ket{2}$  \\
    $\ket{-_{1}}$ & $\ket{1}$   & $\ket{-_{1}}\otimes\ket{0}$  \\
    $\ket{-_{1}}$ & $\ket{2}$   & $\ket{-_{1}}\otimes\ket{1}$  \\
    \hline
    $\ket{-_{2}}$ & $\ket{0}$   & $\ket{-_{2}}\otimes\ket{1}$  \\
    $\ket{-_{2}}$ & $\ket{1}$   & $\ket{-_{2}}\otimes\ket{2}$  \\
    $\ket{-_{2}}$ & $\ket{2}$   & $\ket{-_{2}}\otimes\ket{0}$  \\
    \hline
  \end{tabular}

\end{center}
\end{table}
As listed in Table \ref{tab:tbl2}, various uniform superposition states can be asserted using the circuit presented in Fig \ref{fig:superpositionAsr}. For instance, if the expected state is $\ket{\psi}=\ket{-_{1}}$, then the ancilla qutrit needs to set to $\ket{1}$. In this configuration, the ancilla qutrit measurement result of $0$ indicates the absence of an assertion error, while a different measurement outcome suggests otherwise.

\section{Evaluation}
\label{sec:Evaluation}
In this section, we present various use cases to assess the efficacy of our proposed assertion circuits.
\subsection{Asserting Classical States}
 We commenced by testing a ternary quantum half-adder (HA) circuit from the prior work \cite{haghparast2017towards}. Our objective is to validate the final results through sample inputs. 
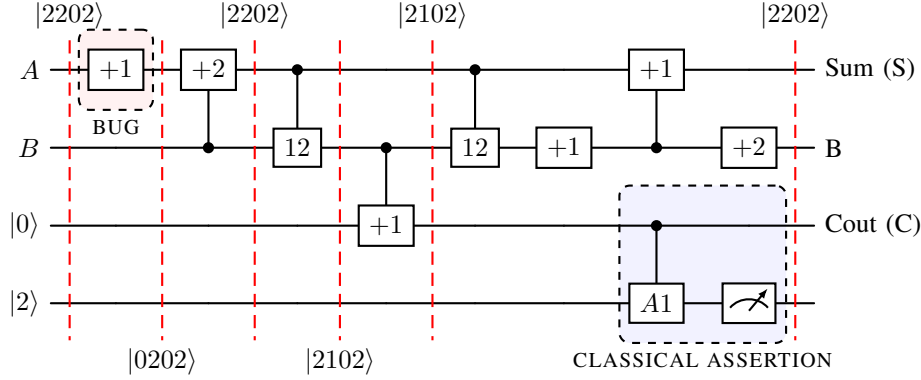
\begin{figure*}[h]
  \centering
  \begin{quantikz}
    \lstick{$A$} \slice{$\ket{2202}$}&\gate[1]{+1}\gategroup[1,steps=1,style={dashed,rounded
corners,fill=red!5, inner
xsep=0.1 pt},background,label style={label
position=below,anchor=north,yshift=-0.2cm}]{{\sc bug}} \slice[label style={label
position=below,yshift=-4.6cm}]{$\ket{0202}$}& \gate[1]{+2} \slice{$\ket{2202}$}& \ctrl{1} && \ctrl{1}& &\gate[1]{+1} &&\rstick{Sum (S)} \\
    \lstick{$B$}& & \ctrl{-1} & \gate[1]{12} \slice[label style={label
position=below,yshift=-4.6cm}]{$\ket{2102}$}& \ctrl{1} &  \gate[1]{12} & \gate[1]{+1} & \ctrl{-1}& \gate[1]{+2}&\rstick{B}\\
     \lstick{$\ket{0}$}& & & & \gate[1]{+1} \slice{$\ket{2102}$}&&&\ctrl{1}\gategroup[2,steps=2,style={dashed,rounded
corners,fill=blue!5, inner
xsep=0.1 pt},background,label style={label
position=below,anchor=north,yshift=-0.2cm}]{{\sc classical assertion}}&& \rstick{Cout (C)}\\ 
    \lstick{$\ket{2}$}&    & &&&& &\gate{A1}& \meter{}\slice{$\ket{2202}$} &
   \end{quantikz}
   \caption{Classical assertion in a half adder circuit for $\ket{1}$ using the proposed circuit in Fig. \ref{fig:classicAsr} with ancilla qutrit being $\ket{2}$. The functionality of the gates, e.g., the '\textit{12}' gate, is shown in Table\ref{tab:tbl1}. Here, a non-zero output will be measured because of the bug, leading to an assertion error (There is no overflow, though it is expected).}
  \label{fig:Classical_testCase}
\end{figure*}
Let us consider a specific scenario where the input states are set as $\ket{AB} = \ket{22}$. As a result, the expected output should be $\ket{SC} = \ket{11}$ with an overflow occurred since the carry-out (Cout) should be equal to one. In this case, we select our assertion circuit and set $\ket{2}$ as its ancilla input to $assert(carryout== \ket{1})$. Upon measuring the outcome of the ancilla qutrit, when the Cout is $\ket{1}$, we obtain $\ket{0}$ without disturbing the original HA circuit. However, should the Cout have a value of $\ket{0}$ or $\ket{2}$, it indicates that the overflow is incorrect. This leads to an assertion error, as the ancilla qutrit is no longer $\ket{0}$.

To illustrate the process of debugging the HA circuit, let's hypothetically consider a scenario where a bug was introduced during the design phase. For instance, let's assume that a [+1] gate was mistakenly used before the controlled-[+2] gate as the initial gate in the HA circuit (as illustrated in Fig.\ref{fig:Classical_testCase}).  
Despite this discrepancy, the initial values remain consistent for both \ket{AB} and the two additional qutrits. However, it is important to note that since the Cout qutrit should be checked 
 
it must be the qutrit that controls the [A1] gate. After integrating the suggested assertion circuit with an ancilla state of $\ket{2}$, an error becomes evident upon measuring the fourth qutrit (the bottom-most qutrit). In this case, an undesired outcome would arise for the Cout. Instead of the intended value of 1, it would be 2. Consequently, with the proposed classical assertion circuit to detect a value of 1, the measurement outcome would yield a non-zero result, leading to an assertion error.

Given that the quantum cost of utilizing each MS gate is one unit, the quantum cost (QC) of the classical assertion circuit is calculated to be 4. The depth of the classical assertion circuit would be 4, reflecting the number of logical levels within the circuit.
\subsection{Asserting Superposition States}
\begin{figure}[htbp]
  \centering
      \begin{quantikz}[
slice all, remove end slices=1,
slice style = {shorten <=-0.1cm,
               shorten >=-0.1cm},
slice label style = {yshift=-26mm}, 
slice titles = $\psi_{\fpeval{\col-1}}$
        ]
        \lstick{$\ket{\psi}$}& \gate{Ch2}\gategroup[1,steps=1,style={dashed,rounded
corners,fill=red!5, inner
xsep=0.1 pt},background,label style={label
position=above,anchor=north,yshift=0.2cm}]{{\sc bug}}&  \gategroup[2,steps=3,style={dashed,rounded
corners,fill=blue!5, inner
xsep=0.1 pt},background,label style={label
position=above,anchor=north,yshift=0.2cm}]{{\sc superposition assertion}}& \gate{A1}& && \\
        \lstick{$\ket{1}$}    &  &\gate{Ch1}& \ctrl{-1} & \gate{Ch2} &\meter{} & 
      \end{quantikz}
  \caption{Checking the uniform superposition generation using the assertion circuit in Fig.  \ref{fig:superpositionAsr} with ancilla qubit being $\ket{1}$.}
  \label{fig:superposition_testCase}
\end{figure}
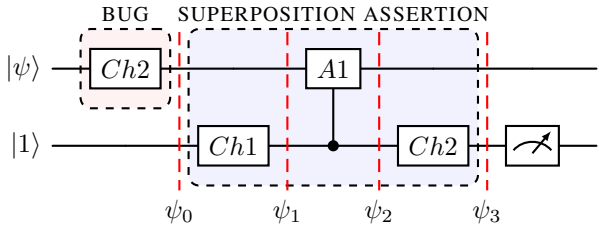

To illustrate the use of the proposed circuit in asserting ternary superposition states, we conducted a test using a reference circuit designed to produce the expected superposition state $\tfrac{1}{\sqrt{3}} [ \ket{0} + \omega\ket{1} + \omega^2\ket{2}]$. This state is the result of applying a [Ch1] gate to the input value $\ket{1}$. In a hypothetical scenario, we inadvertently employed a [Ch2] gate instead of a [Ch1] gate while starting with an initial input state of $\ket{1}$, as shown in Fig.\ref{fig:superposition_testCase}, which led to the creation of a hypothetical bug. As a result, we obtained $\tfrac{1}{\sqrt{3}} [ \ket{0} + \omega^2\ket{1} + \omega\ket{2}]$ instead of the expected state. By using the superposition assertion circuit, we detected this discrepancy, confirming it as an assertion error. According to the Table \ref{tab:tbl2}, as the expected state is $\ket{-_1}$, the ancilla qutrit is initialized to be $\ket{1}$.

\textit{\textbf{proof.}} According to Fig \ref{fig:superposition_testCase}, if \(\ket{\psi}\) is equal to \(\ket{1}\), the circuit produces the following states as a result of the mentioned bug:

\begin{small}
\begin{align*}
\ket{\psi_{0}} &= \tfrac{1}{\sqrt{3}}[\ket{01} + \omega^2\ket{11} + \omega\ket{21}] \\
\ket{\psi_{1}} &= \tfrac{1}{3}[ \ket{0}\otimes(\ket{0}+\omega\ket{1}+\omega^2\ket{2}) +\\
&+\omega^2\ket{1}\otimes(\ket{0}+\omega\ket{1}+\omega^2\ket{2}) + \\
&+\omega\ket{2}\otimes(\ket{0}+\omega\ket{1}+\omega^2\ket{2})]\\
&= \tfrac{1}{3}[(\ket{00}+\omega\ket{01}+\omega^2\ket{02}) +\\
&+(\omega^2\ket{10}+\ket{11}+\omega\ket{12}) +\\
&+(\omega\ket{20}+\omega^2\ket{21}+\ket{22}) ]\\
\ket{\psi_{2}} &= \tfrac{1}{3} [\ket{00} + \omega\ket{11} + \omega^2\ket{22} + \omega^2\ket{10} + \ket{21} + \omega\ket{02} + \omega\ket{20} + \\ 
&+\omega^2\ket{01} + \ket{12}] \\
\ket{\psi_{3}} &= \tfrac{1}{3\sqrt{3}}[\ket{00} + \ket{01} + \ket{02} + \omega\ket{10} + \ket{11} + \omega^2\ket{12} + \omega^2\ket{20} \\ &+ \ket{21} + \omega\ket{22} + \\
&+ \omega^2\ket{10} + \omega^2\ket{11} + \omega^2\ket{12} + \ket{20} + \omega^2\ket{21} + \omega\ket{22} + \omega\ket{00} +\\ &+ \omega^2\ket{01} + \ket{02} + \\
&+ \omega\ket{20} + \omega\ket{21} + \omega\ket{22} + \omega^2\ket{00} + \omega\ket{01} + \ket{02} + \ket{10} +\\ &+ \omega\ket{11} + \omega^2\ket{12}]\\
&=> \\
\ket{\psi_{3}} &= \tfrac{3}{3\sqrt{3}} [\ket{02} + \omega^2\ket{12} + \omega\ket{22}] = \ket{-_2}\otimes\ket{2}
\end{align*}
\end{small}
As it can be seen, we did not obtain a $\ket{0}$ as our output for the second qutrit, which is essential for confirming the correctness of the final result. Since a value other than $\ket{0}$ is measured  (in this case, $\ket{2}$), an assertion error is reported. \\

\subsection{Asserting for Entangled States}
\begin{figure}[htbp]
  \centering
      \begin{quantikz}
        \lstick{$\ket{\alpha}$}& &\gate{Ch1}\gategroup[2,steps=2,style={dashed,rounded
corners,fill= green!5, inner
xsep=0.1 pt},background,label style={label
position=above,anchor=north,yshift=0.2cm}]{{\sc circuit \cite{Corbaci_2016}}}& \ctrl{1} & \ctrl{2}\gategroup[3,steps=2,style={dashed,rounded
corners,fill=blue!5, inner
xsep=0.1 pt},background,label style={label
position=below,anchor=north,yshift=-0.2cm}]{{\sc entangled assertion}} && &\rstick[2]{$\ket{\psi_{\alpha\beta}}$}\\
        \lstick{$\ket{\beta}$}& \gate{+1}\gategroup[1,steps=1,style={dashed,rounded
corners,fill=red!2, inner
xsep=0.1 pt},background,label style={label
position=below,anchor=north,yshift=-0.2cm}]{{\sc bug}}&  & \targ{} & & \ctrl{1}&&\\
        \lstick{$\ket{0}$}    &  &&  & \gate{A1} & \gate{A2}& \meter{}&
      \end{quantikz}
  \caption{Checking entangled qutrits introduced in \cite{Corbaci_2016} using the proposed assertion circuit in Fig. \ref{fig:3b} to detect the bug. The entangled circuit with ancilla qutrit being $\ket{0}$ asserts for $\frac{1}{\sqrt{3}} (\ket{00} + \ket{11} + \ket{22})$.}
  \label{fig:entangled_testCase}
\end{figure}
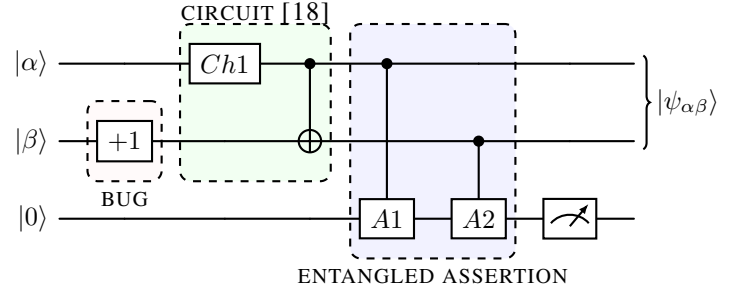

To showcase the efficacy of our proposed circuit for asserting a ternary entangled state, we apply it to another circuit presented in reference \cite{Corbaci_2016} for generating entangled states. The circuit is shown in Fig. \ref{fig:entangled_testCase}. We aim to ascertain the accuracy of the circuit's output and detect any potential bugs. Let us consider a scenario where the input state has changed because a [+1] gate is added mistakenly on the second qutrit. 
$\ket{\alpha\beta}$ is set to $\ket{00}$. In this specific case, the expected output from the circuit should be a\(\ket{00}\) + b\(\ket{11}\) + c\(\ket{22}\), where $a=b=c= \frac{1}{\sqrt{3}}$. To ensure the accuracy of the result, we employ our proposed assertion circuit for checking the expected entangled state.  
Based on the anticipated output $\frac{1}{\sqrt{3}} (\ket{00} + \ket{11} + \ket{22})$, we find that the appropriate circuit for asserting the target state, which belongs to group \textit{b} in Fig.\ref{fig:entangled_groups}, is the second circuit in Fig.\ref{fig:entangled_circuits}(\textit{b}) with the ancilla = $\ket{0}$ as shown in Fig.\ref{fig:entangled_testCase}. 
In the case that the circuit's outcome aligns with the predetermined expected state, the measured value will be  $\ket{0}$. Otherwise, deviations from the expected state will yield measured values of $\ket{1}$ or  $\ket{2}$, both signifying the occurrence of an assertion error. With the abovementioned bug, i.e., the erroneous application of a [+1] gate on the first qutrit, a non-zero value will be measured for the second qutrit, resulting in an assertion error.

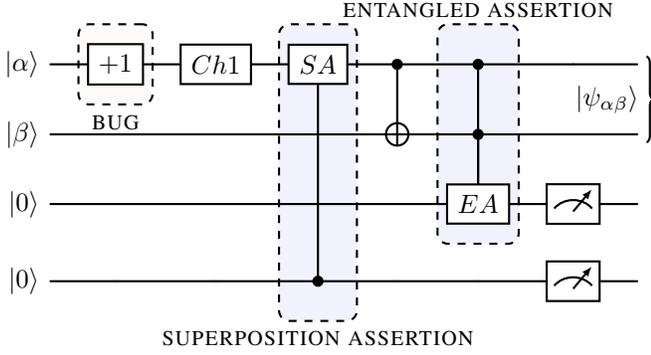
\begin{figure}[htbp]
  \centering
      \begin{quantikz}
        \lstick{$\ket{\alpha}$}&\gate{+1}\gategroup[1,steps=1,style={dashed,rounded
corners,fill=red!2, inner
xsep=0.1 pt},background,label style={label
position=below,anchor=north,yshift=-0.2cm}]{{\sc bug}} &\gate{Ch1}&\gate{SA}\gategroup[4,steps=1,style={dashed,rounded
corners,fill=blue!5, inner
xsep=0.1 pt},background,label style={label
position=below,anchor=north,yshift=-0.2cm}]{{\sc superposition assertion}}& \ctrl{1} & \ctrl{2}\gategroup[3,steps=1,style={dashed,rounded
corners,fill=blue!5, inner
xsep=0.1 pt},background,label style={label
position=above,anchor=north,yshift=0.2cm}]{{\sc entangled assertion}} && \rstick[2]{\hspace{-1.2cm}$\ket{\psi_{\alpha\beta}}$}\\
        \lstick{$\ket{\beta}$}& &  && \targ{} &\ctrl{1} &&\\
        \lstick{$\ket{0}$}    &  &&  && \gate{EA}  & \meter{}&\\
        \lstick{$\ket{0}$}    &  && \ctrl{-3}  &&  & \meter{}&
      \end{quantikz}
  \caption{Asserting entangled qutrits introduced in \cite{Corbaci_2016} using the proposed combined assertion circuits to detect phase errors. The assertion circuit with ancilla qutrits equal to $\ket{00}$ asserts for $\frac{1}{\sqrt{3}} (\ket{00} + \ket{11} + \ket{22})$. The [SA] circuit is the proposed superposition assertion in Fig. \ref{fig:superpositionAsr}, and the [EA] circuit is the proposed entanglement assertion in Fig. \ref{fig:3b}.}
  \label{fig:entangledMergedWithSupperposition_testCase}
\end{figure}

However, our assertion circuit for checking entangled states can only detect changes in classical values and is unable to identify variations in phase characteristics. For instance, the assertion circuit cannot distinguish among $\frac{1}{\sqrt{3}} (\ket{00} + \ket{11} + \ket{22})$ and $\frac{1}{\sqrt{3}} (\ket{00} + \omega\ket{11} + \omega^2\ket{22})$  and $\frac{1}{\sqrt{3}} (\ket{00} + \omega^2\ket{11} + \omega\ket{22})$ states. To remedy this limitation, we propose to employ both the superposition assertion circuit and the entangled assertion circuit, as depicted in Fig. \ref{fig:entangledMergedWithSupperposition_testCase}. 
This way, all the qutrit states listed in Table \ref{tab:tbl3} can be asserted. For example, suppose the anticipated entangled state is $\frac{1}{\sqrt{3}} (\ket{00} + \ket{11} + \ket{22})$. However, when utilizing the original circuit \cite{Corbaci_2016} with the considered bug, i.e., a [+1] gate on the first qutrit before the [Ch1] gate on the same qutrit, the output will be $\frac{1}{\sqrt{3}} (\ket{00} + \omega\ket{11} + \omega^2\ket{22})$, with distinct phases. In order to discern this distinction, we added our proposed superposition assertion circuit, denoted as the SA block. Because the [Ch1] gate's output will not be $\frac{1}{\sqrt{3}} (\ket{00} + \ket{11} + \ket{22})$ with the same phases, the [SA] circuit alters the state of the fourth ancilla qutrit, resulting in a state other than $\ket{0}$ (here, $\ket{2}$). Hence, even though the entangled asserting circuit, which found no errors, keeps its initial input intact, the third and fourth ancilla qutrits' outputs will be in the $\ket{01}$ state because [SA] found an asserting error. Given that $\ket{00}$ is the sole correct assertion, any output other than $\ket{00}$ for the last two qutrits is categorized as an assertion error. To assert and debug the other scenarios, only the ancilla qutrits in Fig. \ref{fig:entangledMergedWithSupperposition_testCase}. need to be adjusted according to the expected stated. This information is provided in both Table \ref{tab:tblx} and Table \ref{tab:tbl2}. The quantum cost of the merged technique is 14.
The evaluation information is summarized in Table \ref{tab:evalSummary}. With this table, one can determine the logical delay and quantum cost associated with the use of each quantum ternary asserting circuit.

\section{Conclusion}
\label{sec:Conclusion}
This paper presents our proposed circuits for assertions in quantum ternary circuits. The supported assertions include classical states, a set of entangled states, and uniform superposition states. With our proposed designs, we show that it is feasible to support dynamic assertions in quantum ternary logic, although they may be more conceptually complex than their binary counterpart. We then provide use cases to show how such assertions can be used to capture bugs in ternary quantum logic. 
\begin{table}[H]
  \begin{center}
  \caption{Table of the test case circuit results and the ancilla qutrits initializing for asserting them based on Fig. \ref{fig:entangledMergedWithSupperposition_testCase}. (Fourth qutrit is the bottom-most) }
  \label{tab:tbl3}
  \centering
  \small 
  \noindent\begin{tabular}{|c|c|c|c|}
    \hline
    &&\multicolumn{2}{c|}{\textbf{Measurement}} \\
    &&\multicolumn{2}{c|}{\textbf{Output}} \\
    \hline
    Input $\ket{\alpha\beta}$ & Output of& Third  & Fourth \\ \cite{Corbaci_2016}& circuit \cite{Corbaci_2016}&qutrit&qutrit \\
    \hline
    $\ket{00}$ &$\frac{1}{\sqrt{3}} (\ket{00} + \ket{11} + \ket{22})$   &0&0 \\
    $\ket{01}$ &$\frac{1}{\sqrt{3}} (\ket{01} + \ket{12} + \ket{20})$   &1&0 \\
    $\ket{02}$ &$\frac{1}{\sqrt{3}} (\ket{02} + \ket{10} + \ket{21})$   &2&0 \\
    $\ket{10}$ &$\frac{1}{\sqrt{3}} (\ket{00} + \omega\ket{11} + \omega^2\ket{22})$   &0&2 \\
    $\ket{11}$ &$\frac{1}{\sqrt{3}} (\ket{01} + \omega\ket{12} + \omega^2\ket{20})$   &1&2 \\
    $\ket{12}$ &$\frac{1}{\sqrt{3}} (\ket{02} + \omega\ket{10} + \omega^2\ket{21})$   &2&2 \\
    $\ket{20}$ &$\frac{1}{\sqrt{3}} (\ket{00} + \omega^2\ket{11} + \omega\ket{22})$   &0&1 \\
    $\ket{21}$ &$\frac{1}{\sqrt{3}} (\ket{01} + \omega^2\ket{12} + \omega\ket{20})$   &1&1 \\
    $\ket{22}$ &$\frac{1}{\sqrt{3}} (\ket{02} + \omega^2\ket{10} + \omega\ket{21})$   &2&1 \\
    \hline
  \end{tabular}
        
  \end{center}
\end{table}
\begin{table}[H]
  \begin{center}
  \caption{Evaluation summary for each quantum ternary asserting circuit}
  \label{tab:evalSummary}
  \centering
  \small 
  \noindent\begin{tabular}{| c | c | c |}
    \hline
    Proposed Circuit & Cost  & Delay  \\ 
    \hline
    Fig. \ref{fig:classicAsr}. & 4 & 4 \\
    Fig. \ref{fig:entangled_circuits}. & 8 & 7\\
    Fig. \ref{fig:superpositionAsr}. & 6 & 6\\
    Fig. \ref{fig:entangledMergedWithSupperposition_testCase}. & 8+6 & 7+6\\
    \hline
  \end{tabular}
        
  \end{center}
\end{table}

\bibliographystyle{IEEEtran}
\bibliography{main}

\vfill

\end{document}